\documentclass[%
 reprint,
superscriptaddress,
 amsmath,amssymb,
 aps,
]{revtex4-2}

\usepackage{graphicx}
\usepackage{dcolumn}
\usepackage{bm}

\begin{document}

\preprint{APS/123-QED}

\title{Imaging nanomechanical vibrations and manipulating parametric mode coupling via scanning microwave microscopy}
\author{Hao Xu}
\affiliation{Univ. Lille, CNRS, Centrale Lille, Univ. Polytechnique Hauts-de-France, UMR 8520 - IEMN, F-59000 Lille, France }
\author{Srisaran Venkatachalam}
\affiliation{Univ. Lille, CNRS, Centrale Lille, Univ. Polytechnique Hauts-de-France, UMR 8520 - IEMN, F-59000 Lille, France }
\author{Toky-Harrison Rabenimanana}
\affiliation{Univ. Lille, CNRS, Centrale Lille, Univ. Polytechnique Hauts-de-France, UMR 8520 - IEMN, F-59000 Lille, France }
\author{Christophe Boyaval}
\affiliation{Univ. Lille, CNRS, Centrale Lille, Univ. Polytechnique Hauts-de-France, UMR 8520 - IEMN, F-59000 Lille, France }
\author{Sophie Eliet}
\affiliation{Univ. Lille, CNRS, Centrale Lille, Univ. Polytechnique Hauts-de-France, UMR 8520 - IEMN, F-59000 Lille, France }
\author{Flavie Braud}
\affiliation{Univ. Lille, CNRS, Centrale Lille, Univ. Polytechnique Hauts-de-France, UMR 8520 - IEMN, F-59000 Lille, France }
\author{Eddy Collin}
\affiliation{Univ. Grenoble Alpes, Institut NEEL - CNRS UPR2940, 25 rue des Martyrs, BP 166, 38042 Grenoble Cedex 9, France}
\author{Didier Theron}
\affiliation{Univ. Lille, CNRS, Centrale Lille, Univ. Polytechnique Hauts-de-France, UMR 8520 - IEMN, F-59000 Lille, France }
\author{Xin Zhou}
\email{corresponding author:  xin.zhou@cnrs.fr or xin.zhou@iemn.fr}
\affiliation{Univ. Lille, CNRS, Centrale Lille, Univ. Polytechnique Hauts-de-France, UMR 8520 - IEMN, F-59000 Lille, France }

\date{\today}

\begin{abstract}
In this study, we present a novel platform based on scanning microwave microscopy for manipulating and detecting tiny vibrations of nanoelectromechanical resonators using a single metallic tip. The tip is placed on the top of a grounded silicon nitride membrane, acting as a movable top gate of the coupled resonator. We demonstrate its ability to map mechanical modes and investigate mechanical damping effects in a capacitive coupling scheme, based on its spatial resolution. We also manipulate the energy transfer coherently between the mode of the scanning tip and the underlying silicon nitride membrane, via parametric coupling. Typical features of optomechanics, such as anti-damping and electromechanically induced transparency, have been observed. Since the microwave optomechanical technology is fully compatible with quantum electronics and very low temperature conditions, it should provide a powerful tool for studying phonon tunnelling between two spatially separated vibrating elements, which could potentially be applied to quantum sensing.  
\end{abstract}

\maketitle


Micro- and nano-electromechanical systems (MEMS and NEMS) have been of great interest for several decades due to their relevance in various applications and fundamental research \cite{lyshevski2018mems, bachtold2022mesoscopic}. Their features, including small scale, high resonance frequency, and low dissipation, make them favorable for different sensing applications, such as mass, force, and photon sensing \cite{spletzer2006ultrasensitive, spletzer2008highly, yang2006zeptogram, chaste2012nanomechanical, zhou2019chip}. These applications have also been accompanied by research motivations to understand the intrinsic and extrinsic sources of dissipation in NEMS/MEMS \cite{imboden2014dissipation, unterreithmeier2010damping, schmid2011damping, barois2012ohmic}. The intrinsic nonlinearity and tunability of their mechanical modes are ideal tools for modelling other systems. For example, studies of the coherent energy exchange in the coupled mechanical modes reveal the classical features of St$\Ddot{u}$ckelberg interferometry \cite{seitner2016classical} and optomechanically induced transparency \cite{okamoto2013coherent, pokharel2022coupling}. Recently, advances in optomechanical technologies have made it possible to use mechanical resonators for quantum sensing and quantum engineering \cite{cattiaux2021macroscopic, mercier2021quantum}. The impressive achievements and advances, mentioned above, keep on driving researchers to develop novel configurations with special features for reading and manipulating tiny nanomechanical motions. 

To date, various sensitive transduction schemes have been achieved for detecting and manipulating nanomechanical resonators. These mainly include methods based on electrical detection or optical interferometry \cite{kaynak2023atmospheric, ti2022dynamics, miller2019spatially}. The former technology is widely used for testing electrically integrated nanomechanical systems, which requires the design and nanofabrication of on-chip circuits to detect and control vibrating elements. However, this technique does not have the spatial resolution required to study localised mechanical vibrations. On the other hand, optical interferometry offers high sensitivity in terms of spatial resolution and direct transmission of the detected signals. Unfortunately, mechanical resonators made of high stiffness materials with transparent or low reflective properties are difficult to detect directly at low light intensities. However, high intensities are detrimental for cryogenic applications. Both probe microscopy \cite{garcia2008imaging, halg2021membrane, rieger2014energy, garcia2007mechanical} and electron microscopy \cite{barois2012ohmic, tsioutsios2017real} have also been developed, based on the decreasing size of the vibrating elements. They not only facilitate the readout of these mechanical resonators with extremely small interaction volumes (such as carbon nanotubes), but also open up access to the study of local mechanical properties \cite{barois2012ohmic}, engineering mechanical impedance \cite{rieger2014energy}, and hybridised nanomechanical systems \cite{halg2021membrane}. However, the scanning tip acts as an active sensing/probing element and relies on mechanical interaction with samples, and additional complex driving circuits on the chip, as reported in previous studies on the probe microscopy \cite{zhou2014scanning, garcia2008imaging}. As well, electron microscopy needs an additional force source to excite the nano-vibration of the massive element, which poses a challenge for combinations with standard electron microscopy chambers. 

Scanning microwave microscopy (SMM) can detect capacitances at the aF scale and their variations, allowing for the characterisation of a wide range of materials \cite{huber2010calibrated, gu2017broadband}. It has the dual advantage of highly sensitive microwave technology and the spatial resolution of the scanning tip. In recent years, microwave interferometry has been used to detect tiny displacements of MEMS, enabling the transmission of drive and detection signals through a single gate electrode. \cite{zhou2021high, pokharel2022coupling, venkatachalam2023effects}. Microwave techniques are directly compatible with today's quantum electronics and microwave photons are much less energetic, making them suitable for cryogenic environments \cite{bernier2017nonreciprocal, mercier2019realization}. These previous works inspired and motivated us to extend the SMM approach by integrating it with microwave interferometry to detect and manipulate the nanomechanical resonator through a single scanning tip for the first time. 

In the present work, we demonstrate the ability of the SMM to spatially image nanomechanical vibration modes, without requiring scanning tip contact with the vibrating silicon nitride membrane. It enables the investigation of spatial dependencies (over the membrane size) while scanning the tip position. Coherent energy transfer between the scanning tip and the membrane resonators via parametric coupling was investigated based on phonon-cavity nanoelectromechanics \cite{pokharel2022coupling}. Typical optomechanical features were observed in this phonon-cavity electromechanics, including ``optomechanical damping effects" and electromechanically induced transparency. Our study extends the current applications of SMM and provides a novel platform for the detection of nanomechanical systems integrated into a complex circuit \cite{zhou2019chip}. Furthermore, it may serve as a valuable tool for investigating phonon-phonon interactions where precise control of the interactions is required \cite{fong2019phonon, xu2022non}.

\begin{figure*}
  \includegraphics[width=0.90\textwidth]{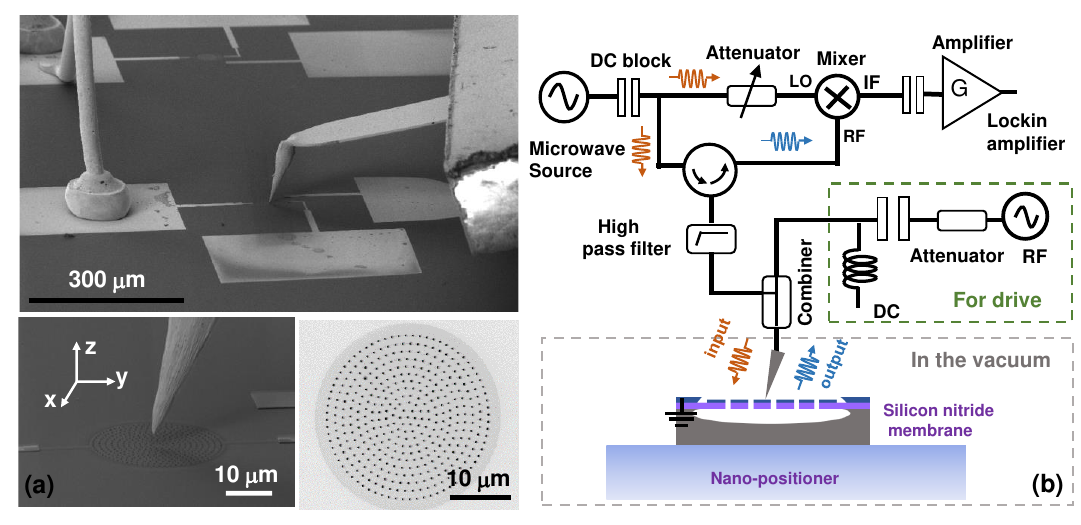}
  \caption{(a) On the top, SEM image of a metallic tip suspended at the top of the silicon nitride circular membrane which is covered with an Al thin film \cite{zhou2021high}. The tip is glued on a printed circuit board (PCB). The membrane is grounded through  bonding wires. The zoomed-in SEM image is shown at the bottom left. The silicon nitride circular membrane, released from the silicon substrate through a two-step dry etching process (see SI), is presented at the bottom right \cite{zhou2021high, xu2023fabrication}. (b) Schematic diagram of the high frequency electrical setup for driving and detection of the chip. The membrane is placed on the top of a nano-positioner and held in a vacuum chamber during measurement (below $\sim 5\times10^{-4}$ mbar) at room temperature. Details of the setup and tip information are presented in SI.}
  \label{sch:setup}
\end{figure*}
$\textbf{Measurement setup of SMM}$ A metallic tip is used as a suspended and movable top gate, in this measurement. It was soldered onto a 50 $\Omega$ impedance PCB. The nanoelectromechanical resonator, measured in this work, is a silicon nitride circular membrane, with a thickness of $\approx$ 90 nm and radius of $a\approx$ 16 $\mu$m, as shown in Fig.\ref{sch:setup}(a). Its top is covered with a 20 nm aluminium layer. Fabrication details can be found in our previous report \cite{zhou2021high}. The sample is fixed on a nano-positioner with $\textit{X-Y-Z}$ relative displacements up to a nanometer precision. The tip, used in this setup, is a commercial scanning probe (25PT300D), having a resonance frequency $\Omega_{tip}/(2 \pi)$ = 15.39 kHz with a linewidth $\gamma_{tip}/(2\pi)$ = 30 Hz. The tip is not perpendicular to the sample surface and forms a small angle with the Z axis. The setup with the tip and the sample is located in the vacuum chamber of a scanning electron microscope (SEM). 

%
%
The high-frequency setup is schematically depicted in Fig. \ref{sch:setup} (b). To excite mechanical vibrations, an electrostatic force $F$ is generated between the tip and the coupled membrane at the frequency $\Omega_d$, $F = \frac{\partial C_g(\mu)}{\partial \mu}V_{dc}V_{ac}\cdot Cos(\Omega_d t)$, through the combination of a $\textit{dc}$ voltage $V_{dc}$ and a $\textit{ac}$ signal $V_{ac}\cdot Cos(\Omega_d t)$. Here, $\mu$ is the mechanical displacement, corresponding to either the tip or membrane. The $ C_g(\mu)$ is the coupling capacitance between the tip and membrane. The readout of the mechanical motion relies on the microwave interferometry \cite{zhou2021high}. A microwave signal at the frequency $\omega /2\pi$ = 6 GHz is directly shined to the membrane through the same tip. The $C_g(\mu)$ can be written as $C_{g0}(1-\frac{\mu}{H})$, where $H$ is the distance between the tip and the membrane, and the $C_{g0}$ is the capacitance value independent from $\mu$. The interaction between $\mu$ and the input microwave signal produces a reflected signal at the frequency $\omega + \Omega_d$ (see SI), which is then read out by a lock-in amplifier after frequency down-conversion. Both the driving and the detection signals pass through the single tip, avoiding the complexities of wiring the chip. This setup is universal and adapts to any NEMS driven by electrostatic forces in a capacitive coupling scheme with small stray capacitance values. 
\begin{figure*}
  \includegraphics[width=0.98\textwidth]{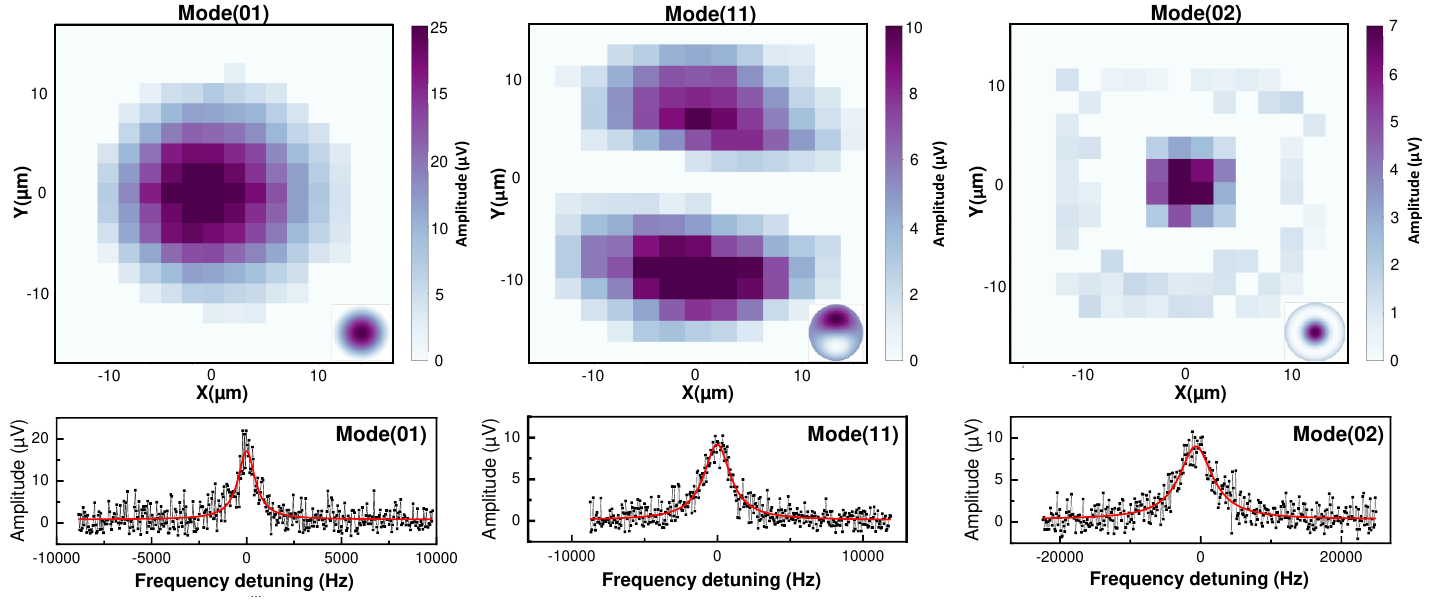}
  \caption{Top: measured amplitude of the resonance peaks as a function of the spatial position (X-Y), corresponding to each vibration mode. Inset, calculation results of the expected modes based on solutions of the n-th Bessel function of the first kind \cite{zhou2021high}. Bottom: Examples of the silicon nitride membrane mechanical response, measured around the location having the maximum signal amplitude in the X-Y plane.}
  \label{sch:modeMap}
\end{figure*}

$\textbf{Measurement results and discussions}$ The procedure of mapping mechanical mode shapes begins with positioning the tip relative to the membrane. The SEM is used to help the alignment of the tip and the membrane, as shown in Fig. \ref{sch:setup}. After calibrating the tip position, the tip is then positioned at a fixed height $H$ above the top of the membrane. Compared with previous work of detecting graphene nanodrums using optical interferometry \cite{davidovikj2016visualizing}, we focused on exciting only three modes of this membrane and measured resonance frequencies $\Omega_{0,1}/(2\pi) \approx$ 8.82 MHz, $\Omega_{1,1}/(2\pi) \approx$ 14.06 MHz, and $\Omega_{0,2}/(2\pi) \approx$ 20.25 MHz, respectively. The vibration mode was mapped by sweeping the tip position in the X-Y plane with a step of 2 $\mu$m while exciting the the linear mechanical responses of the membrane with constant external $ac$ and $dc$ signals. The spatial mode maps depicting the vibrations of the circular membrane are generated by plotting the amplitude of the detected signal at the resonance frequency as a function of the tip position in the X-Y plane, as shown in Fig. \ref{sch:modeMap}. These measurement results reflect the relative vibration amplitudes of the nanomechanical membrane, as the amplitude of the detected electrical signal is directly proportional to the mechanical vibration amplitude \cite{zhou2021high}. The measurement results reveal a minor spatial asymmetry compared with the calculated results shown in the inset of Fig.\ref{sch:modeMap}. This is due to the imperfect alignment between the tip and the Z-axis of the chip.

\begin{figure}
  \includegraphics[width=0.49\textwidth]{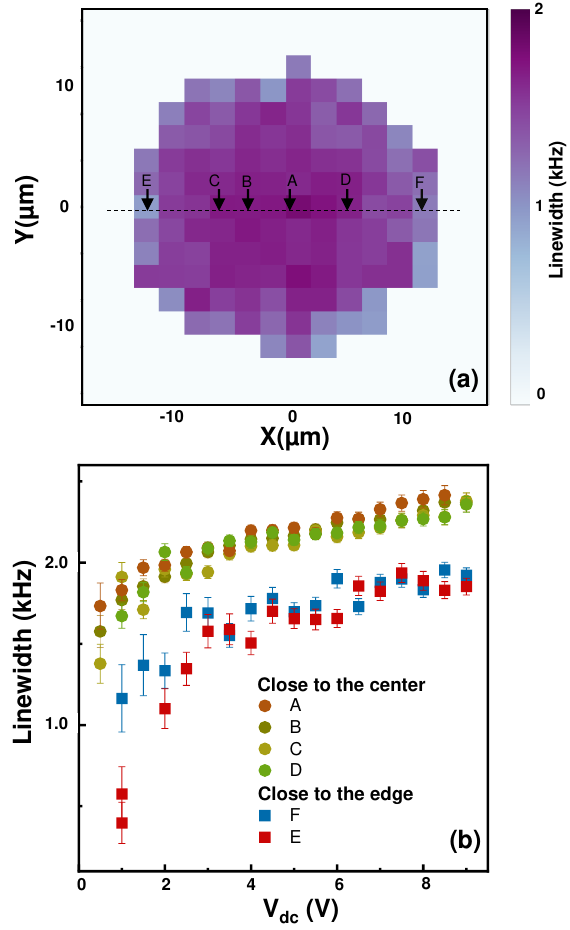}
  \caption{(a) Linewidth map for the fundamental mode in the spatial position (X-Y), measured at $V_{dc}$ = 4 V, $V_{ac}$ = 40 m${V_p}$, and H = 550 nm. (b) Linewidth of the mechanical response as a function of $V_{dc}$, corresponding to the tip position from A to F marked on the map (a), along the X axis. The measurement is performed with $V_{ac}$ = 55 m${V_p}$, and H = 450 nm. Position A is taken as the reference point. Position B is offset from A by $\approx$ 4 $\mu$m, and both C and D are offset by $\approx$ 5 $\mu$m in two opposite directions. Positions E and F are offset by 12 $\mu$m and 10 $\mu$m respectively.}
  \label{sch:damp_pos}
\end{figure}

The scanning tip provides a unique opportunity to spatially image the mechanical properties of the nanoelectromechanical resonator. Figure \ref{sch:damp_pos} (a) shows the linewidth map of the first mode. It is constructed by Lorentzian function fitting of the linear mechanical responses at each position in the X-Y plane corresponding to the mode map presented in Fig. \ref{sch:modeMap}. The damping rate at the centre of the membrane has higher values than the damping rate near clamped edges. This is contrary to the common intuition that the oscillating element closer to the clamping edge suffers more from so-called clamping loss \cite{tsaturyan2017ultracoherent}. To further verify this phenomenon, we then remeasure the linewidth of the resonance at 6 different positions along the diameter of the circular membrane, from A to F as marked on \ref{sch:damp_pos} (a). The difference in linewidth cannot be neglected, as shown in \ref{sch:damp_pos} (b). A monotonous increase with $V_{dc}$ is observed, with approximately the same dependence all over the surface. This shall be discussed below; but it is clear that the linewidth corresponding to $V_{dc}$ around 1 V is about 50 $\%$ of the value measured with $V_{dc} >$ 7 V. The value of the linewidth measured near the edge seems to drop rather sharply. 

As mentioned above, the excited mechanical motion relies on the interaction between the tip and the membrane. The electric fields are symmetrically distributed across the X-Y plane when the tip is close to the centre of the membrane. Conversely, when the tip is close to the edge of the membrane, there may be less effective electric field lines terminating at the membrane due to the reduction in the effective coupling area. Consequently, the coupling strength at the tip position near the center of the membrane is significantly different from that when the tip is near the edge. The effective back actions from the tip on the membrane become smaller when the tip approaches the edges of the membrane, even when the biased $V_{dc}$ and $V_{ac}$ at the tip are held as constant values during the measurement. This could result in less dissipation from mechanical bending in the vibrating membrane, leading to smaller values of the damping rate. To better understand this phenomenon, a more complex mechanical model is required. The model should consider the effects of non-uniform driving forces on mechanical deformation, which have not been fully explored in previous reports \cite{unterreithmeier2010damping,schmid2011damping,  sementilli2022nanomechanical}.
\begin{figure}
  \includegraphics[width=0.40\textwidth]{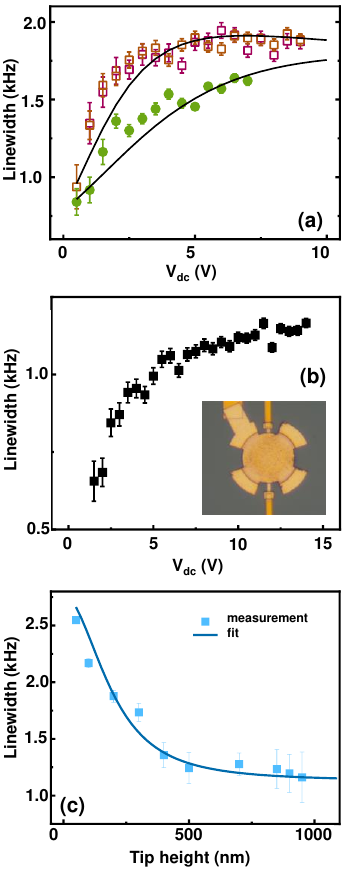}
  \caption{(a) Linewidth of the fundamental mode as a function of $V_{dc}$ bias, measured at the centre of the membrane with $V_{ac}$ = 30 mV$_p$, and H = 300 nm for the square with brown and purple colored edges, and H = 400 nm for the green dots. Black lines: theory fit using Eq.\ref{eqn:dampingModel} with $V_{ac}$ = 20 mV, $\gamma_0/2\pi$ = 785 Hz, and $\sigma_0$ = 17.5 $\mu$S. (b) Linewidth of the silicon nitride membrane nanomechanical resonator capacitively coupled to a suspended Al membrane top gate and an optical image of the corresponding device \cite{zhou2021high}. The device is measured at $V_{ac}$ = 1 mV$_p$. (c) Linewidth of the membrane as a function of the tip height, measured at $V_{dc}$ = 5 V and $V_{ac}$  =  30 mV$_p$. The solid dark blue line is a fit result.}
  \label{sch:damp_vdc}
\end{figure}
Besides, the damping rates present a $V_{dc}$ dependence, presented in Fig.\ref{sch:damp_pos} (b). We reduce the tip height and carefully measure the linewidth versus $V_{dc}$ of the second drum sample, which has the same radius and is located on the same chip. The $\gamma_m$ increases with increasing $V_{dc}$ and begins to saturate slightly at high $dc$ voltages, as shown in Fig. \ref{sch:damp_vdc} (a). The linewidth nearly doubles within 10 V. It is quite different from previous reports that damping exhibited a quadratic behavior \cite{rieger2012frequency, barois2012ohmic}. To verify whether the tip induces the $dc$ voltage dependence of the mechanical damping in the membrane, we examined another membrane device with a different gate coupling scheme. In this device design, the silicon nitride membrane is capacitively coupled to an Al membrane top gate clamped with four feet on the substrate \cite{zhou2021high}, instead of the scanning tip, as shown in the Fig.\ref{sch:damp_vdc} (b). The measurement results of the linewidth exhibit the same tendency as that measured with the tip. It indicates that the $V_{dc}$ dependent damping rate originates from the nanoelectromechanical membrane in the capacitive transduction scheme.  
\\
We start from the definition of the quality factor based on the capacitive circuit damping model, $Q$ = $2\pi ( \frac{1}{2} m_{eff} \cdot \Omega^2_m \cdot \mu_0^2 ) / (R \int_0^{2 \pi / \Omega_{m}} I^2 \mathrm{d} t)$ with the mechanical displacement $\mu(t) = \mu_0 \cdot Sin(\Omega_m t)$, to understand this damping behavior \cite{imboden2014dissipation}. The term in the parentheses corresponds to the kinetic energy stored in the vibrating membrane (over a period $2\pi/ \Omega_m$) and the term in the denominator is the dissipated energy in the whole circuit. The $m_{eff}$ corresponds to the effective mass integrated over the mode shape \cite{hauer2013general}. In the measurement shown in Fig.\ref{sch:damp_vdc}, the tip position is fixed in the X-Y plane and the mode shape is the same for all $V_{dc}$ dependence measurements. So, the $m_{eff}$, here, is a constant value for the mode. We used the capacitive circuit damping model and obtain the equivalent mechanical resistance $R_m = \gamma_0 \cdot m_{eff} / (V_{dc} \partial C_g / \partial \mu )^2$ through an analog of the $LCR$ circuit \cite{imboden2014dissipation, lazarus2010simple}, where $C_g(H, \mu)$ is the coupling capacitance between the tip and the membrane at height $H$. Here, the $\gamma_0$ is the intrinsic mechanical damping rate caused by clamping losses and materials. The total dissipation resistance is $R$ = $(R_m^{-1}+ \sigma_0)^{-1}$, where $\sigma_0$ is the conductance of the highly resistive substrate in parallel with mechanical resistance. The displacement current $I$ is therefore given by $I \approx (V_{dc}+V_{ac}) \frac{C_{g0}}{H}\mu_0 \Omega_m \cdot Cos(\Omega_m t) + C_{g0} \Omega_m  V_{ac} \cdot Cos(\Omega_m t)$, with the approximation of $\partial C_g / \partial \mu \approx C_{g0} /H$ and $V_{dc} \gg V_{ac}$. The $C_{g0}(H)$ is the initial capacitance excluding contributions from the mechanical displacement $\mu(t)$. Thus, the expression of damping is presented as

\begin{equation}
\begin{aligned}
  \gamma_m & \approx \gamma_0 + \frac{(V^2_{dc}+2V_{ac}V_{dc} \frac{H}{\mu_0}) C^2_{g0}}{m_{eff} H^2} \cdot \frac{1}{\sigma_0 (1 + \epsilon V^2_{dc})}, \\
  \epsilon & \approx \frac{1}{ \sigma_0 \gamma_0 \cdot m_{eff}} (\frac{C_{g0}}{H})^2.
\end{aligned}
\label{eqn:dampingModel}
\end{equation}
A similar model has also been developed in graphene or carbon nanotube based nanomechanical systems that have non-negligible resistive elements in their components to understand the damping issue \cite{song2012stamp, barois2012ohmic}. As demonstrated by this model, the $\gamma_m$ shows a high dependence on the coupling capacitance, thus making it sensitive to the $H$. Using the unique capability of the SMM, this property has been proved by measurement results of the resonance linewidth as a function of the tip height, as shown in Fig.\ref{sch:damp_vdc} (c). The solid lines presented in Fig.\ref{sch:damp_vdc} are the best fit results by using the Eq.\ref{eqn:dampingModel}, by taking parameters in a reasonable range, such as $m_{eff}\approx$ 4$\times 10^{-14}$ kg, $C_{g0} \cdot H \approx$ 5.56$\times 10^{-22} $ F$\cdot$m, $V_{ac}$ = 18 mV, $V_{dc}$ = 5 V, $\gamma_0/2\pi$ = 1130 Hz, and $\sigma_0$ = 17.5 $\mu$S. This $\sigma_0$ value is consistent with the high resistivity properties of the silicon substrate beneath the membrane. Further studies of the model and more careful calibrations are still needed to reach a perfect quantitative fit.

\begin{figure}
  \includegraphics[width=0.40\textwidth]{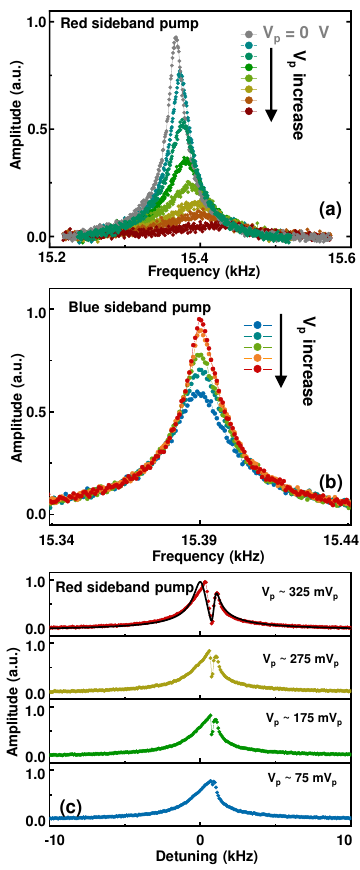}
  \caption{(a) Mechanical responses of the tip. Red sideband pumping of the membrane, measured with a probe signal $V_{ac}[\Omega_d]$ = 5 mV$_p$ sweeping around $\Omega_{tip}$. Pump amplitude $V_{ac}[\Omega_p]$ varies from 75 to 375 mV$_p$ with a step increase of 50 mV$_p$, from dark green to dark red color. The grey curve corresponds to the situation of the no pump tone. (b) Mechanical response of the tip, measured by pumping the membrane at its blue sideband and probing the tip around the frequency $\Omega_{tip}$ with an amplitude $V_{ac}[\Omega_d]$ = 5 mV$_p$. The pump amplitudes $V_{ac}[\Omega_p]$ = 100, 150, 175, 200, 210  mV$_p$, from blue to red color. (c) Red sideband pumping of the membrane and weak probing of it around its resonance frequency $\Omega_{01}$ with $V_{ac}[\Omega_d]$ = 30 mV$_p$ and $V_{ac}[\Omega_p]$  =  325, 275, 175, and 75 mV$_p$ respectively. Here, all measurements are performed through biasing $dc$ voltage $V_{dc}$ = 5 V, with a tip height $H \approx$ 50 nm. The black line is a calculation result based on Eq.\ref{eqn:analog}. In these measurements, a background is subtracted from each measured frequency response curve.}
  \label{sch:doubletone}
\end{figure}

The integration of a scanning tip and microwave interferometry also provides an approach to manipulate the coherent energy transfer between the tip and the underlying membrane through parametric mode coupling. By analogy with the optomechanical system \cite{pokharel2022coupling, kumar2023microwave, weis2010optomechanically}, we choose the silicon nitride membrane resonator as the phonon cavity and pump its sideband at the frequency $\Omega_p$ = $\Omega_{01} \pm \Omega_{tip}$, where ``+" and ``-" correspond to the so-called blue and red sideband pumping, respectively. This sideband pumping technique can be used to observe the ``optomechanical damping effect". The effective linewidth of the tip is modulated by the phonon-cavity force, described by $\gamma_{eff}$ = $\gamma_{tip} \pm |\gamma_{opt}|$. The $|\gamma_{opt}|$ is the optomechanical damping whose value is proportional to the pump amplitude $V_{p}$ and the ``+" (``-") corresponds to red (blue) sideband pumping scheme \cite{aspelmeyer2014cavity, zhou2021electric, pokharel2022coupling}. To observe it, we probe the tip with a very weak signal amplitude sweeping around $\Omega_{tip}$ and pump strongly the phonon-cavity at its red or blue sideband. Figure \ref{sch:doubletone} (a) shows that the $\gamma_{tip}$ increases with increasing the $V_{p}$, for the red sideband pumping scheme. The resonance of the tip is almost damped out at $V_p$ = 375 mV$_p$. The energy of the tip, in the form of phonons, is effectively pumped out. The variations in the $\gamma_{tip}$ as a function of the $V_p$ follow the theoretical descriptions of a phonon-cavity nanomechanical system \cite{pokharel2022coupling}. This technique has been used to dynamically reduce the thermal occupation of nanomechanical resonators to approach quantum ground states in optomechanical systems \cite{teufelGround}. For the blue sideband, as shown in Fig.\ref{sch:doubletone} (b), the linewidth decreases as the pump amplitude increases because the phonon occupation of the tip increases. It may provide an opportunity to improve the sensitivity of a conventional scanning tip, by profiting from the blue sideband pumping technique. 

In addition, nanoelectromechanically induced transparency can be observed in a red sideband pumping scheme, when one of the coupled modes is driven with a large signal amplitude. Its principle relies on using the sideband pump to create destructive interference with the probe signal, demonstrating the potential application of signal filters. In our previous work \cite{pokharel2022coupling}, a theoretical model was developed to describe this classical nature of the induced transparency by analogy with the optomechanical system, without considering the Duffing nonlinearity. When the membrane is pumped at $\Omega_p$ = $\Omega_{01} -\Omega_{tip}$, the displacement of the probed membrane is given by 

\begin{equation}
    \begin{aligned}
      \mu(\Omega_d) &= \frac{1}{2m_{eff}  \Omega_{01}} \cdot \frac{C_{g0}V_{dc} V_{ac}[\Omega_d]}{H} \\
      \cdot &\frac{1}{(-(\Omega_d - \Omega_{01}) - i \frac{\gamma_m}{2}) - \frac{n_{p} g_0^2}{-(\Omega_d - \Omega_{01})-i\frac{\gamma_{tip}}{2}}},
    \end{aligned}
\label{eqn:analog}
\end{equation}
where $n_p$ is the phonon number generated by the pumping force and $g_0$ is the single phonon coupling strength. The destructive interference mechanism is presented in the denominator part. The pump and probe tones generate a signal at the frequency $\Omega_d-\Omega_p$ around the $\Omega_{tip}$ through frequency down-conversion. This signal then is projected back to the membrane due to the pump, forming a destructive interference with the probe signal. This is the classical nature of the induced transparency in this two-tone measurement \cite{pokharel2022coupling}. To measure it, we drive the phonon-cavity at $V_{ac}[\Omega_d]$ = 30 mV$_p$ and pump its red sideband with different pump amplitudes. This drive amplitude is critical to produce adequate phonons cycling between the photon-cavity and the coupled tip. Figure \ref{sch:doubletone}(c) shows clear dips in the resonances of the membrane, which exhibit slightly non-linear behaviours due to large driving forces. As the pump amplitude increases, more phonons are projected back to coherently cancel the drive signal, resulting in a deeper dip. These measurement results indicate that this tip-membrane coupled system can be used for studying phonon tunneling and could therefore serve for quantum sensing \cite{fong2019phonon, xu2022non, chan2001quantum, xu2022observation}.

\textbf{Conclusion} We have presented a novel platform based on SMM in which a scanning tip, acting as a suspended top gate, is used to drive and detect the tiny displacements of a nanomechanical resonator. The scanning tip, with its features of spatial resolution, has demonstrated its unique ability to image the mechanical vibration mode and  investigate the linear damping properties of the membrane. In addition, anti-damping effects and electromechanically induced transparency have also been demonstrated in this parametrically coupled tip-membrane system. This platform extends the current applications of SMM to the MEMS/NEMS domain and provides a unique opportunity to study electromechanical properties at the nanoscale. Specifically, it allows for the convenient study of vibrating elements embedded in complex circuits, such as microwave optomechanical circuits  \cite{zhou2019chip}. The concept of this experimental configuration will further facilitate research activities to go beyond the current frontiers of quantum sensing and quantum engineering. For example, it can be used to study a few numbers of phonon tunneling and non-reciprocal phonon transfer across vacuum through quantum fluctuations \cite{fong2019phonon, xu2022non, chan2001quantum, xu2022observation}, by taking advantage of microwave readout schemes which feature high sensitivity and low heating effects. 

\textbf{Author Contributions}: X.Z. conceived the design of the experiment, built the measurement setup, performed the high-frequency measurements, and developed the analytical model. S.V. and X.Z fabricated the measured samples. D.T. and H.X. contribute to manually operating the tip. T.H participated in plotting data. C.B., S.E., and F.B. provide technical support for sample setup. X.Z. wrote the manuscript in consultation with E.C.. All authors have scientific discussions. X.Z. supervised the project.

\textbf{Acknowledge}: The authors would like to acknowledge financial support from the French National Research Agency, ANR-MORETOME, No. ANR-22-CE24-0020-01, and Chist-ERA NOEMIA project with contract ANR-22-CHR4-0006-01. This work was partly supported by the French Renatech network. X.Z. would like to thank Gianluca Rastelli, Eva Weig, Pierre Vielot, and Anthony Ayari for fruitful discussions.

\bibliography{apssamp}

\end{document}